\pdfoutput=1

\documentclass[11pt,twoside,a4paper,cmspaper,final,collab]{cms-tdr}
\def\svnVersion{166730}\def\cmsCernNo{2012-110}\def\cmsCernDate{\today}\def\cmsMessage{Submitted to the Journal of High Energy Physics}

\begin{document}\cmsNoteHeader{SUS-11-013}

\hyphenation{had-ron-i-za-tion}
\hyphenation{cal-or-i-me-ter}
\hyphenation{de-vices}

\RCS$Revision: 127376 $
\RCS$HeadURL: svn+ssh://svn.cern.ch/reps/tdr2/papers/SUS-11-013/trunk/SUS-11-013.tex $
\RCS$Id: SUS-11-013.tex 127376 2012-06-04 13:37:50Z mwalker $
\newcommand{\jt}{\ensuremath{H_{\mathrm{T}}}\xspace}
\newcommand{\GeVC}{\ensuremath{\GeV\!/c}}
\newcommand{\gevc}{\ensuremath{\GeV\!/c}}
\newcommand {\abseta} {|\eta|}
\newcommand{\mzero}{\ensuremath{{m_0}}}
\newcommand{\azero}{\ensuremath{A_{0}}}
\newcommand{\mhalf}{\ensuremath{m_{1/2}}}
\newcommand{\tb}{\ensuremath{\tan\beta}}
\newcommand{\nall}{\widetilde{{\chi}}^{0}}
\newcommand{\call}{\widetilde{{\chi}}^\pm}
\newcommand{\ntwo}{\widetilde{{\chi}}^{0}_{2}}
\newcommand{\cone}{\widetilde{{\chi}}^\pm_{1}}
\newcommand{\none}{\widetilde{{\chi}}^{0}_{1}}
\newcommand{\nunit}[2]{#1\,\mbox{#2}}
\newcommand{\met}{\ensuremath{E_{\mathrm{T}}^{\text{miss}}}\xspace}
\newcommand{\curlumi}{714}
\providecommand{\FIXME}[1]{({\bf FIXME: #1})}
\providecommand{\HERWIGPP} {{\textsc{herwig++}}\xspace}
\providecommand{\JIMMY} {{\textsc{jimmy}}\xspace}
\providecommand{\NLOJETPP} {{\textsc{nlojet++}}\xspace}
\providecommand{\Et}{E_{\mathrm{T}}}
\providecommand{\met}{\mbox{${\hbox{$\vec{E}$\kern-0.5em\lower-.1ex\hbox{/}}}_T~$}}
\providecommand{\MET}{\mbox{${\hbox{$E$\kern-0.5em\lower-.1ex\hbox{/}}}_T~$}}
\providecommand{\pthat}{\ensuremath{\hat{\text{p}}_\mathrm{T}}\xspace}
\providecommand{\kthat}{\ensuremath{\hat{\text{k}}_\mathrm{T}}\xspace}
\providecommand{\xt}{\ensuremath{\text{x}_\mathrm{T}}\xspace}
\providecommand{\ptjet}{\ensuremath{p_{\mathrm{T,jet}}}\xspace}
\providecommand{\FASTNLO} {{fast\textsc{nlo}}\xspace}

\providecommand{\rbthm}{\rule[-2ex]{0ex}{5ex}}
\providecommand{\rbthr}{\rule[-1.7ex]{0ex}{5ex}}
\providecommand{\rbtrm}{\rule[-2ex]{0ex}{5ex}}
\providecommand{\rbtrr}{\rule[-0.8ex]{0ex}{3.2ex}}
\providecommand{\relmet}{\ensuremath{\text{MET}/\sum{E_{\mathrm{T}}}}}
\providecommand{\LvStartup}{\ensuremath{\mathcal{L}=\text{10}^\text{31}\,\text{cm}^\text{$-$2}\,\text{s}^\text{$-$1}}\xspace}

\newcommand{\fullLumi}{4.98\fbinv}
\newcommand{\procLumi}{\ensuremath{4.98\fbinv}}
\newcommand{\iso}{\text{iso}}
\newcommand{\noniso}{\text{noniso}}
\newcommand{\eiso}{\epsilon_\iso}
\newcommand{\eisomu}{\eiso(\mu)}
\newcommand{\eisotr}{\eiso(\text{track})}
\newcommand{\esb}{\epsilon_\mathrm{sb}}
\newcommand{\TTbar}{\ttbar}
\newcommand{\ST}{\ensuremath{S_\mathrm{T}}}
\newcommand{\Jpsi}{\ensuremath{{J /\psi}}}

\cmsNoteHeader{SUS-11-013} 
\title{Search for anomalous production of multilepton events in pp collisions at $\sqrt{s} = 7\TeV$}

\date{\today}

\abstract{
A search for anomalous production of events with three or more isolated leptons in pp collisions at $\sqrt{s} = 7\TeV$ is presented.
The data, corresponding to an integrated luminosity of \procLumi, were collected by the CMS experiment at the LHC during the 2011 run. The search is applicable to any model of new physics that enhances multiple lepton production. The observed multilepton events are categorized into exclusive search channels based on the identity and kinematics of the objects in the events. An estimate of the standard-model background rates from data is emphasized, but simulation is also used to estimate some of the background rates.  The search results are interpreted in the context of supersymmetry, including both $R$-parity-conserving and $R$-parity-violating models. We derive exclusion limits as a function of squark, gluino, and chargino masses.
}
\hypersetup{%
pdfauthor={CMS Collaboration},%
pdftitle={Search for anomalous production of multilepton events in pp collisions at sqrt(s) = 7 TeV},%
pdfsubject={CMS},%
pdfkeywords={CMS, physics}}

\maketitle 

\section{Introduction}
Events with three or more prompt leptons are rarely produced by standard-model (SM) processes in proton-proton collisions. It is therefore possible that physics processes beyond the standard model (BSM) at the LHC may first be observed in multilepton final states. In this article, we describe a search for anomalous production of multilepton events based on data collected with the Compact Muon Solenoid (CMS) experiment at the LHC. The analysis described here is similar in structure to the search described in Reference~\cite{Chatrchyan:2011ff}, but uses a substantially larger integrated luminosity of \procLumi~\cite{CMS-PAS-SMP-12-008}.

Although this search is not tailored to any particular model, it is well-suited for constraining models that enhance multilepton production.  Certain scenarios in supersymmetry (SUSY) satisfy this requirement. Supersymmetry is a well-known candidate for a BSM theory that solves the hierarchy problem, allows for the unification of the gauge couplings, and may provide a candidate particle to solve the dark matter problem~\cite{Nilles:1983ge,Haber:1984rc,deBoer:1994dg,Baer:1991xs,Baer:1994nr,Baer:1995va}.

In SUSY, $R$-parity is defined as $R_\mathrm{p} = (-1)^{3B+L+2s}$, where $B$ and $L$ are the baryon and lepton numbers and $s$ is the particle spin~\cite{1978PhLB.76.575F}. All SM particles have $R_\mathrm{p} = +1$ while all superpartners have $R_\mathrm{p} = -1$.  In models where $R$-parity is conserved, superpartners can only be produced in pairs, and the lightest supersymmetric particle (LSP) is stable and a candidate dark matter particle. In addition, $R$-parity conservation ensures proton stability. We study scenarios with either the neutralino or the gravitino as the LSP. We also compare models in which the LSP is stable to $R$-parity violating (RPV) cases in which the LSP decays to SM particles.

If the gravitino is the LSP, one of the sleptons, a lepton superpartner, can be the next-to-lightest supersymmetric particle (NLSP). Scenarios of this type arise in a wide class of theories of gauge-mediated supersymmetry breaking (GMSB)~\cite{Dimopoulos:1996vz,Dimopoulos:1996yq,Culbertson:2000am}.  Multilepton final states arise naturally in the subset of the GMSB parameter space where the right-handed sleptons are flavor degenerate, the so-called ``slepton co-NLSP scenario''~\cite{Dimopoulos:1996yq,Culbertson:2000am,Ruderman:2010kj,Alves:2011wf}.

In $R$-parity conserving models, the stable, weakly-interacting LSPs appear to produce momentum imbalance, which is measured using \met, the net transverse energy (\ET) carried away by undetected particles. In addition, the decays of massive squarks and gluinos lead to large total jet momentum, which is measured using \jt, the scalar sum of the \pt of all reconstructed jets. These features make \jt and \met good observables for discriminating $R$-parity conserving models from the SM.

In contrast, the lack of a stable LSP in RPV models makes \met a poor discriminator, which motivates a second type of analysis using \ST, which we define as the scalar sum of \met, \jt, and the \pt of all isolated leptons. The value of \ST~reflects the sum of the parent particle masses if most of the energy is reconstructed as leptons, jets, or \met. Therefore, signal events generated by new heavy particles are expected to have much larger values of \ST~than SM backgrounds.

To have sensitivity to both types of models, we apply two separate sets of selections to the data. The first classifies the data into two regions of \jt, $\jt < 200\GeV$ and $\jt > 200\GeV$, and two regions of \met, $\met < 50\GeV$ and $\met > 50\GeV$. The second classifies events into three regions of \ST: low ($\ST<300\GeV$), medium ($300\GeV<\ST < 600\GeV$), and high ($\ST >600\GeV$).

In both cases, we further categorize events with either three or four isolated leptons (electrons, muons, or taus) based on the number of opposite-sign same-flavor (OSSF) electron or muon pairs and whether the event contains an OSSF pair with its invariant mass in the Z mass region, between 75 and 105\GeV~\cite{Chatrchyan:2011ff}.

\section{Detector and Event Trigger}

The central feature of the CMS apparatus is a superconducting solenoid, of 6~m internal diameter, providing a field of 3.8\unit{T}. Within the field volume are a silicon pixel and strip tracker, a crystal electromagnetic calorimeter (ECAL) and a brass/scintillator hadron calorimeter. Muons are measured in gas-ionization detectors embedded in the steel return yoke. Extensive forward calorimetry complements the coverage provided by the barrel and endcap detectors. A more detailed description can be found in Ref.~\cite{:2008zzk}. Data from pp interactions must satisfy the requirements of a two-level trigger system. The first level performs a fast selection for physics objects (jets, muons, electrons, and photons) above certain thresholds.  The second level performs a full event reconstruction.

CMS uses a right-handed coordinate system, with the origin at the nominal interaction point, the $x$-axis pointing to the centre of the LHC, the $y$-axis pointing up (perpendicular to the LHC plane), and the $z$-axis along the counterclockwise-beam direction. The polar angle, $\theta$, is measured from the positive $z$-axis and the azimuthal angle, $\phi$, is measured in the $x$-$y$ plane. The pseudorapidity, $\eta$, is a transformation of the polar angle defined by $\eta = -\ln (\tan (\theta/2))$.

Data for this search are collected with single-electron, double-electron, single-muon, and double-muon triggers, as well as an electron-muon (e-$\mu$) trigger.  We use jet triggers, which are based on the summed jet $\pt$ in the event, for ancillary purposes in this analysis. The thresholds on the triggers varied over the course of 2011 to cope with the increasing instantaneous luminosity of the LHC. To establish uniformity across the sample, we apply an offline threshold corresponding to the most stringent used during the run.

To ensure that the trigger efficiency is high and stable for our selected events, we require each event selected by the dielectron (dimuon) triggers to have at least one electron (muon) with $\pt > 20\GeV$ and another with $\pt > 10\GeV$. In the case of the e-$\mu$ trigger, we require the leading lepton to have $\pt > 20\GeV$ and the next-to-leading lepton to have $\pt > 10\GeV$. Single-lepton-triggered events are required to have either a muon with  $\pt > 35\GeV$ or an electron with $\pt > 85\GeV$.

Triggers based on \jt are used only for determining electron and muon trigger efficiencies. We estimate the efficiency of the single-electron (single-muon) triggers by selecting isolated electrons (muons) in the \jt-triggered dataset.

A single electron (muon) above the $\pt$ specified threshold has a trigger efficiency of $94.5\%\pm0.7$\% ($87.0\%\pm0.5$\%). Dilepton triggers have an efficiency of $99.0^{+1.0}_{-2.0}$\% for dielectrons,  $92.6\%\pm2.5\%$ for dimuons, and $96.9\%\pm2.0$\% for the e-$\mu$ trigger. The uncertainties in the efficiencies are largely due to the low number of dilepton events in the $H_\mathrm{T}$-triggered datasets.

\section{Object Selection}
We consider events that contain electrons, muons, and taus. In this analysis, additional leptons (besides those used to satisfy trigger requirements) are required to have $\pt \ge 8\GeV$ and $|\eta| < 2.1$. Details of the reconstruction and identification can be found in Ref.~\cite{EGM-10-004} for electrons and in Ref.~\cite{MUO-10-002} for muons.

Electrons are reconstructed as electromagnetic showers in the ECAL that are associated with a track. Electrons must satisfy $\Delta\phi$ (where $\phi$ is the azimuthal angle) and $\Delta\eta$ requirements for matching between the shower and the track, and the track must satisfy criteria designed to remove photon conversions in the detector material. Muons are required to have matching tracks in the tracking and muon detectors and to be consistent with a minimum-ionizing particle in the calorimeters.

Tau leptons can decay either leptonically ($\tau_\ell$) to electrons or muons, or hadronically ($\tau_\mathrm{h}$). Electrons and muons arising from tau lepton decays are selected as described above. The hadronic decays yield either a single track (one-prong) or three tracks (three-prong), occasionally with additional electromagnetic energy from neutral pion decays, and are reconstructed using the hadron plus strips algorithm~\cite{2012JInst.7.1001C}. In this analysis, we focus on one-prong $\tau_\mathrm{h}$ decays, which have a much lower background.

To ensure that the electrons, muons, and taus are isolated, track $\pt$ and calorimeter-tower $\ET$ values are summed in a cone of $\Delta R < 0.3$ (0.4 for electrons) around the object, where $\Delta R = \sqrt{\Delta\phi^2 + \Delta\eta^2}$ is the distance in the $\eta-\phi$ plane. This sum is divided by the object's $\pt$. The resulting ratio $I_\text{rel}$ is required to be less than 0.15, which selects electrons, muons, and one-prong $\tau_\mathrm{h}$ decays without additional neutral pions.

To be sensitive to one-prong $\tau_\mathrm{h}$ decays accompanied by neutral pions, we reconstruct neutral pions within a cone of $\Delta R < 0.1$ around the isolated track and require the invariant mass of the track and neutral pions to be consistent with that expected from $\tau_\mathrm{h}$ decay.  We use the CMS particle-flow (PF) algorithm~\cite{PFT-10-004,PFT-08-001} to identify the neutral pions and to calculate the visible $\pt$ of the $\tau_\mathrm{h}$ candidate. A requirement on the isolation is imposed as before; however, since neutral pions deposit energy near the charged track, the calorimeter tower $\ET$ is summed in a cone of $0.1 < \Delta R < 0.3 $ around the isolated track. In this case, the ratio $I_\text{rel}$ is the isolation energy divided by the sum of $\pt$ of the track and neutral pions and is required to be less than 0.15. All electrons, muons, isolated tracks, and isolated tracks with neutral pions are required to point to within 1\unit{cm} of the primary vertex and to be separated by at least $\Delta R > 0.3$ from other particles in the event.

To estimate the efficiencies of the electron and muon identification and isolation requirements, we use the method described in Ref.~\cite{ref:tag-and-probe} for Z\,$\to\ell^+\ell^-$ events. The simulation models the efficiencies correctly to within 2\% ($1$\%) for electrons (muons). We verify that the simulation accurately models the efficiencies for isolated tracks and isolated tracks with neutral pions by comparing the number of Z\,$\to \tau_\mu \tau_\mathrm{h}$ events in the simulation to the number found in the data. We measure the ratio of the efficiency in data and simulation for single-prong $\tau_\mathrm{h}$ events to be $1.02 \pm 0.04$.

Jets are reconstructed with the anti-$k_T$ clustering algorithm~\cite{Cacciari:2008gp}, using a distance parameter of 0.5. The jet reconstruction is based on PF objects. They can include leptons. Jets are required to have $|\eta| < 2.5$ and $\pt > 40\GeV$ and to be distant by $\Delta R > 0.3$ from any isolated electron, muon, or track.

Events with an OSSF pair mass below 12\GeV are rejected to exclude events with \JPsi\ mesons, $\Upsilon$ mesons, low-mass Drell--Yan processes, and photon conversions.

\section{Background Processes and Systematic Uncertainties}

Several SM processes can produce signatures that mimic BSM events with three or more leptons. The largest background remaining after requiring three leptons originates from the production of Drell--Yan pairs (including Z boson production) in association with jet activity, in which the a third, fake lepton is produced from a jet or a photon. The probability for a jet to produce an isolated-lepton candidate depends on the type of jet, the jet and lepton $\pt$ spectra, the number of pile-up interactions (additional proton-proton collisions in the same beam crossing), and the number of jets in the event. These factors may be inaccurately modeled in the simulation; therefore, we estimate the background from jets using dilepton and jet-enriched data samples as follows.

We measure the number of isolated electron or muon background events to be the product of the number of isolated ($K^{\pm}$ or $\pi^{\pm}$) tracks in the dilepton sample and two fractions. The two fractions are: i) the number of nonisolated leptons divided by the number of nonisolated tracks in the dilepton sample, and ii) the ``isolation efficiency ratio'' which is the ratio of the probability for a lepton originating from a jet to pass the lepton isolation requirement to the probability for a track candidate to do so.

A complication worthy of note is the dependence of the isolation efficiency ratio on the relative abundance of charm and bottom quarks, which differs between the QCD and dilepton samples. Therefore, we parametrize the efficiency ratio as a function of the impact parameter distribution of non-isolated tracks for various QCD samples and then choose the efficiency ratio value that corresponds to the measured impact parameter distribution of the dilepton sample.  We measure the ratio of the number of isolated leptons to isolated tracks from jets in dilepton data , i.e., the product of ratios (i) and (ii) above, to be $1.34\%\pm0.35$\% ($1.45\%\pm0.15$\%) for electrons (muons).  (Contributions from dileptonic decays of $\TTbar$ are subtracted throughout.)  The dominant source of systematic uncertainty in the ratio measurements is the difference in jet properties of the dilepton and jet-dominated QCD samples.

To understand the backgrounds in channels with $\tau_\mathrm{h}$, we extrapolate the isolation sideband $0.2<I_\text{rel}<0.5$ to the signal region $I_\text{rel}<0.15$.  The ratio of the number of isolated tracks in the two regions is $15\% \pm 3\%$. We study the variation of this ratio for a number of jet-dominated samples and assign a 30\% systematic uncertainty to account for the differences measured between the samples and for the variation in the results from using different functional forms to parameterize the distribution.  We use the ratio to calculate the contribution of jets mimicking taus in the three- and four-lepton samples by applying it to the number of two-lepton events with tracks.

Some SM background contributions cannot be estimated using data-driven techniques. We perform a detailed simulation of Z/$\gamma^{*}$ + jets, $\TTbar$ quark pairs, double-vector boson production (VV), $\TTbar$V + jets, and WWW + jets using the {\sc MadGraph}~\cite{Maltoni:2002qb} event generator, and of multijet events described by quantum chromodynamics (QCD) using the {\sc Pythia} 8.1 generator~\cite{Sjostrand:2007gs}. We use CTEQ6.6 parton distribution functions (PDF)~\cite{PhysRevD.78.013004}. Next-to-leading order (NLO) cross sections are determined using \textsc{MadGraph}~\cite{Alwall:2011uj}. The detector response is modeled with GEANT4~\cite{Agostinelli:2002hh}.

\begin{figure}[tbh]
\centering
\begin{tabular}{cc}
\includegraphics*[height=5.25cm]{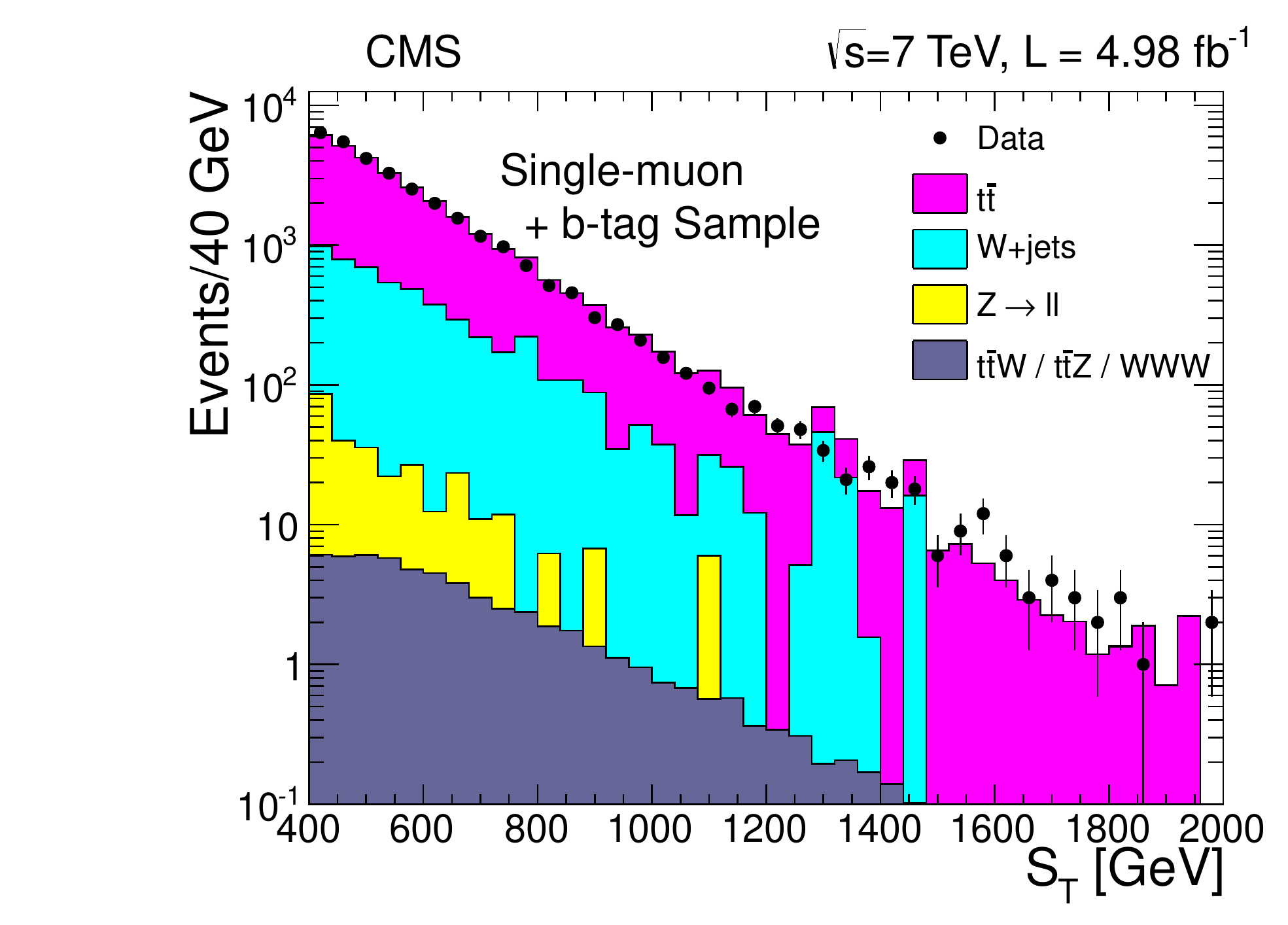}
\includegraphics*[height=5.25cm]{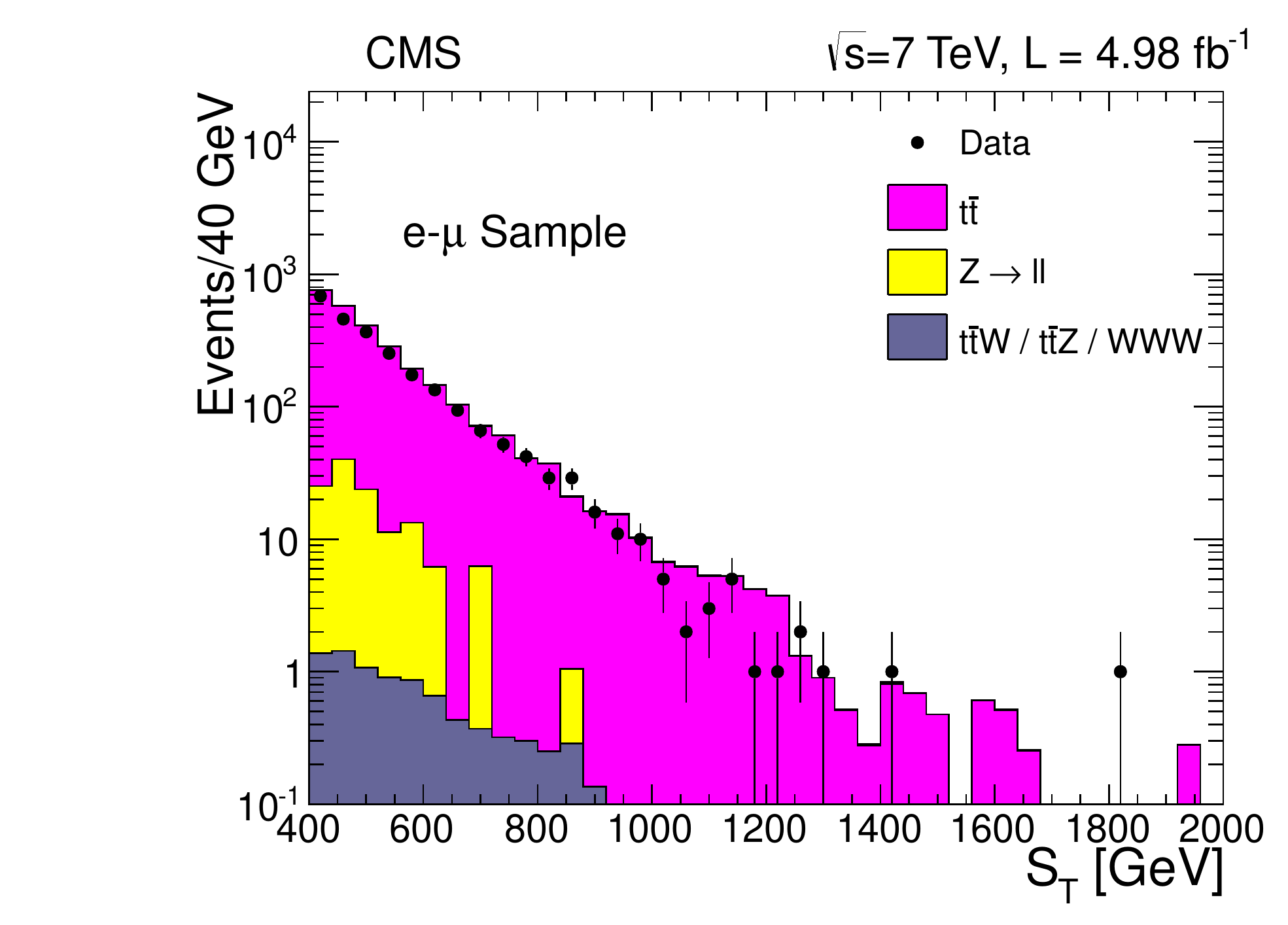}
\end{tabular}
\caption{Comparison of \ST~distributions from data and simulation for two datasets dominated by $\TTbar$: a single-muon with a b-jet sample (left) and an opposite-sign electron-muon sample (right).}
\label{fig:ttbar_st_dist}
\end{figure}

We find the simulation to be adequate for estimating backgrounds from ZZ\,$\to 4\ell$, W$^{\pm}$Z\,$\to 3\ell$,
and $\TTbar$+jets\,$\to 2\ell$. To demonstrate the adequacy of the $\TTbar$ simulation, we compare data and simulation for distributions relevant to this process. An example is the \ST~distribution for two control datasets: a dataset with an isolated muon and a jet originating from a bottom quark, and a dataset with an isolated muon and an opposite-sign, isolated electron. These datasets are dominated by $\TTbar$ events for large \ST.  A good agreement between data and simulation is observed (Fig.~\ref{fig:ttbar_st_dist}). The uncertainty on the $\TTbar$ background estimate of 50\% contributes a large systematic uncertainty in channels where this process is prominent. The size of this uncertainty is governed by the limited number of events in the top-enriched control sample used to measure the isolation distribution of muons and electrons from b jets in data. 

There are two different types of photon conversions that give rise to backgrounds in multilepton analyses. The first type is an external conversion in which a real photon produced in the collision interacts with detector material and produces a $\ell^+\ell^-$ pair (usually an e$^+$e$^-$ pair and very rarely a $\mu^+\mu^-$ pair). The electron identification requirements strongly suppress external conversions. The second type of conversion is an internal photon conversion, where the photon is virtual and does not interact with the detector. Internal photon conversions can produce muons almost as often as electrons and can occur in any process that produces photons. If one of the leptons takes most of the photon energy while its partner is very low $\pt$ and not measured, the process is called an asymmetric conversion. When coupled with additional lepton production, this process can be a significant source of background.

Internal conversions may not be properly described in the simulation because of low-energy cutoffs for emitted leptons in the generator; instead, we use data to estimate this background. We assume that the rate for SM processes to produce real photons is proportional to the rate for producing virtual photons that yield asymmetric conversions. This assumption is justified in the leading-logarithm approximation because the virtual photon mass spectrum is strongly peaked in the low mass region, which means virtual photons have kinematics that are very similar to real photons.
The conversion rate for producing a signal lepton via radiation is the ratio of the probability for a photon to produce a valid lepton candidate via asymmetric conversion to the probability for a real photon to pass all of the selection criteria.

\begin{figure}[htbp]
\centering
\begin{tabular}{cc}
\includegraphics*[width=7.5cm]{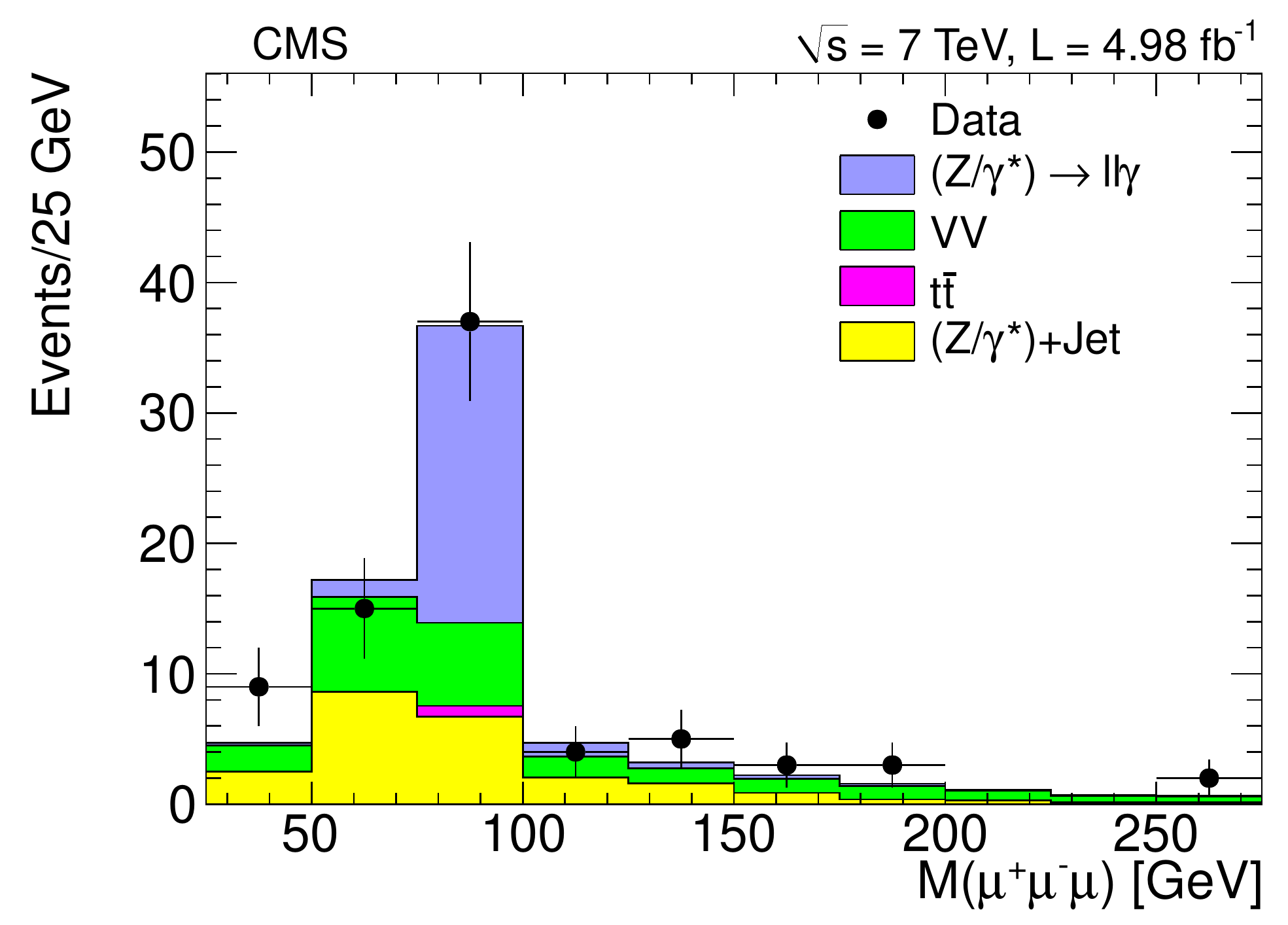}
\includegraphics*[width=7.5cm]{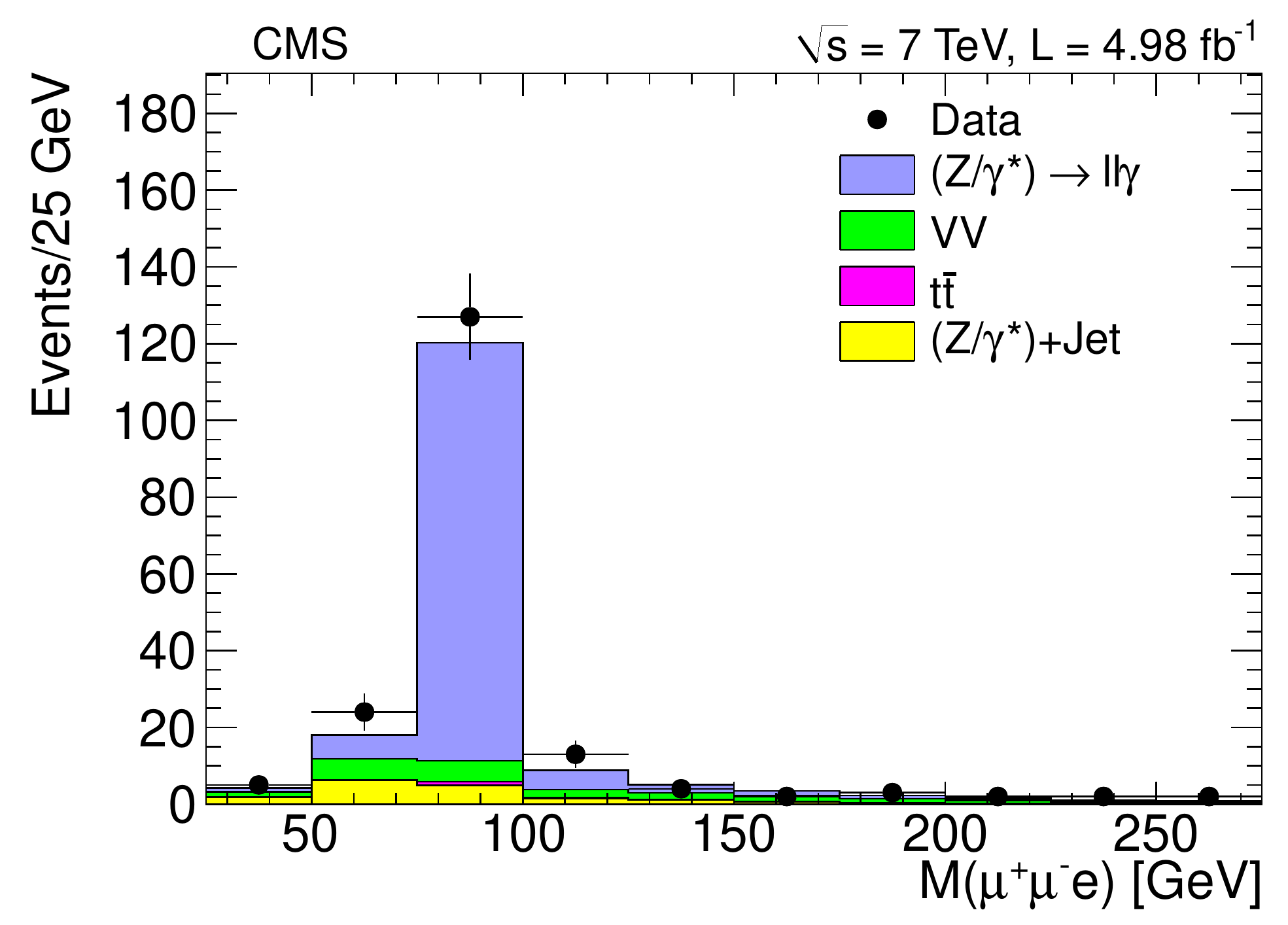}
\end{tabular}

   \caption{Invariant mass distributions of $3\mu$ (left) and $\mu\mu e$ (right) events showing clear Z peaks
caused by the asymmetric-conversion background.}
   \label{fig:fsr}
\end{figure}

For this analysis, the most important source of photon-conversion background involves Z bosons decaying to leptons, and an asymmetric internal conversion of a $\gamma^*$ from one of the leptons.  The radiation of the $\gamma^*$ (virtual photon) moves the mass of the dileptons outside of the Z mass window and through asymmetric conversion the $\gamma^*$ is reconstructed as an additional lepton in the event. We select clean examples of events with final-state radiation by examining three-body masses near the Z peak in channels with both electrons and muons (Fig.~\ref{fig:fsr}).

The ratio of the number of $\ell^+\ell^-\ell^{\pm}$ to $\ell^+\ell^-\gamma$ (real photon) events on the Z peak gives a conversion factor for muons ($C_{\mu}$) of $0.32\%\pm0.08\%\pm0.32\%$ and for electrons ($C_{\Pe}$) of ($1.45\%\pm0.14\%\pm1.45\%$), where the first uncertainty is statistical and the second is systematic. We assign a systematic uncertainty of 100\% to these conversion factors from our underlying assumption that the number of isolated photons is proportional to the number of leptons from asymmetric internal and external conversions because there are insufficient data in control regions to probe this assumption.  We use these conversion factors to estimate the background coming from asymmetric conversions. 

We assign a systematic uncertainty of 2.2\% to the luminosity measurement~\cite{CMS-PAS-SMP-12-008}, which is correlated among all signal channels and the background estimates that are scaled from simulations.
Uncertainties on lepton-identification and trigger efficiencies also contribute to the systematic uncertainty of the result.

\section{Results}
We present the number of observed events and the expectation from SM processes in the \jt and \met regions listed in Table~\ref{tab:ra7}, and in the \ST~regions listed in Table~\ref{tab:rpv}. The rows are labelled by the total number of isolated leptons in the event, the number and mass of OSSF pairs, and the kinematic conditions; the columns indicate how many of those leptons are $\tau_\mathrm{h}$ leptons. 
Reflecting the difficulty of $\tau_\mathrm{h}$ reconstruction, the background increases with the number of $\tau_\mathrm{h}$ leptons.  We do not form OSSF pairs with $\tau_\mathrm{h}$ leptons.

\begin{table}
\small
\begin{center}
\topcaption{Number of observed events summed over electron and muon flavors compared with expectations from simulated and data-driven backgrounds. The labels in the first column refer to whether or not there are OSSF (no-OSSF) pairs, whether $Z\to\ell^+\ell^-$ is excluded (no-Z), and the \jt and \met requirements, which are given in \GeV.  Labels along the top of the table give the number of $\tau_\mathrm{h}$ candidates, 0, 1, or 2. All channels are mutually exclusive. The uncertainties on the expected values include both statistical and systematic uncertainties.
\label{tab:ra7}  }
\begin{tabular}{c|ccc|cc|cc}
\hline
\hline
Selection && \multicolumn{2}{c}{N($\tau_\mathrm{h}$)=0}  & \multicolumn{2}{c}{N($\tau_\mathrm{h}$)=1} & \multicolumn{2}{c}{N($\tau_\mathrm{h}$)=2} \\
  && obs & expected & obs & expected & obs & expected\\
\hline
$4$ Lepton results \\
\hline
$4\ell$ \met$>$50, $\jt>$200,~no Z && 0 & 0.018  $\pm$ 0.005 & 0 & 0.09  $\pm$ 0.06 & 0 & 0.7  $\pm$ 0.7 \\
$4\ell$ \met$>$50, $\jt>200$,~~~~Z && 0 & 0.22  $\pm$ 0.05 & 0 & 0.27  $\pm$ 0.11 & 0 & 0.8  $\pm$ 1.2 \\
$4\ell$ \met$>$50, $\jt<$200,~no Z && 1 & 0.20  $\pm$ 0.07 & 3 & 0.59  $\pm$ 0.17 & 1 & 1.5  $\pm$ 0.6 \\
$4\ell$ \met$>$50, $\jt<$200,~~~~Z && 1 & 0.79  $\pm$ 0.21 & 4 & 2.3  $\pm$ 0.7 & 0 & 1.1  $\pm$ 0.7 \\
$4\ell$ \met$<$50, $\jt>$200,~no Z && 0 & 0.006  $\pm$ 0.001 & 0 & 0.14  $\pm$ 0.08 & 0 & 0.25  $\pm$ 0.07 \\
$4\ell$ \met$<$50, $\jt>$200,~~~~Z && 1 & 0.83  $\pm$ 0.33 & 0 & 0.55  $\pm$ 0.21 & 0 & 1.14  $\pm$ 0.42 \\
$4\ell$ \met$<$50, $\jt<$200,~no Z && 1 & 2.6  $\pm$ 1.1 & 5 & 3.9  $\pm$ 1.2 & 17 & 10.6  $\pm$ 3.2 \\
$4\ell$ \met$<$50, $\jt<$200,~~~~Z && 33 & 37  $\pm$ 15 & 20 & 17.0  $\pm$ 5.2 & 62 & 43  $\pm$ 16 \\
\hline													
$3$ Lepton results \\												
\hline													
$3\ell$ \met$>$50, $\jt>$200,~no-OSSF && 2 & 1.5  $\pm$ 0.5 & 33 & 30.4  $\pm$ 9.7 & 15 & 13.5  $\pm$ 2.6 \\
$3\ell$ \met$>$50, $\jt<$200,~no-OSSF && 7 & 6.6  $\pm$ 2.3 & 159 & 143  $\pm$ 37 & 82 & 106  $\pm$ 16 \\
$3\ell$ \met$<$50, $\jt>$200,~no-OSSF && 1 & 1.2  $\pm$ 0.7 & 16 & 16.9  $\pm$ 4.5 & 18 & 31.9  $\pm$ 4.8 \\
$3\ell$ \met$<$50, $\jt<$200,~no-OSSF && 14 & 11.7  $\pm$ 3.6 & 446 & 356  $\pm$ 55 & 1006 & 1026  $\pm$ 171 \\
$3\ell$ \met$>$50, $\jt>$200,~no Z && 8 & 5.0  $\pm$ 1.3 & 16 & 31.7  $\pm$ 9.6 & -- & -- \\
$3\ell$ \met$>$50, $\jt>$200,~~~~Z && 20 & 18.9  $\pm$ 6.4 & 13 & 24.4  $\pm$ 5.1 & -- & -- \\
$3\ell$ \met$>$50, $\jt<$200,~no Z && 30 & 27.0  $\pm$ 7.6 & 114 & 107  $\pm$ 27 & -- & -- \\
$3\ell$ \met$>$50, $\jt<$200,~~~~Z && 141 & 134  $\pm$ 50 & 107 & 114  $\pm$ 16 & -- & -- \\
$3\ell$ \met$<$50, $\jt>$200,~no Z && 11 & 4.5  $\pm$ 1.5 & 45 & 51.9  $\pm$ 6.2 & -- & -- \\
$3\ell$ \met$<$50, $\jt>$200,~~~~Z && 15 & 19.2  $\pm$ 4.8 & 166 & 244  $\pm$ 24 & -- & -- \\
$3\ell$ \met$<$50, $\jt<$200,~no Z && 123 & 144  $\pm$ 36 & 3721 & 2907  $\pm$ 412 & -- & -- \\
$3\ell$ \met$<$50, $\jt<$200,~~~~Z && 657 & 764  $\pm$ 183 & 17857 & 15519  $\pm$ 2421 & -- & -- \\
\hline			 										
Total $4\ell$  && 37 & 42  $\pm$ 15 & 32.0 & 24.9  $\pm$ 5.4 & 80 & 59  $\pm$ 16 \\
Total $3\ell$  & & 1029 & 1138  $\pm$ 193  & 22693 & 19545  $\pm$ 2457  & 1121 & 1177  $\pm$ 172 \\
Total  && 1066 & 1180  $\pm$ 194 & 22725 & 19570  $\pm$ 2457 & 1201 & 1236  $\pm$ 173 \\

\hline \hline
\end{tabular}
\end{center}

\end{table}

\begin{table}
\small
\begin{center}
\topcaption{
\label{tab:rpv}
Number of observed events summed over electron and muon flavors compared with expectations from simulated and data-driven backgrounds. The labels in the first column refer to how many OSSF pairs there are (OSSF-\#), whether $Z\to\ell^+\ell^-$ is excluded (no-Z), and the \ST~binning. \ST~ranges in \GeV are Low ($\ST<300\GeV$), Mid ($300\GeV< \ST < 600\GeV$), and High ($\ST>600\GeV$). Labels along the top of the table give the number of $\tau_\mathrm{h}$ candidates, 0, 1, or 2. All channels are mutually exclusive. The uncertainties on the expected values include both statistical and systematic uncertainties.}

\begin{tabular}{c|ccc|cc|cc}

\hline\hline
Selection && \multicolumn{2}{c}{N($\tau_\mathrm{h}$)=0}  & \multicolumn{2}{c}{N($\tau_\mathrm{h}$)=1} & \multicolumn{2}{c}{N($\tau_\mathrm{h}$)=2} \\
  && obs & expected & obs & expected & obs & expected\\
\hline
$4$ Lepton results \\
\hline
$4\ell$ (OSSF-0) $S_T$(High) && 0 & 0.0010  $\pm$ 0.0009 & 0 & 0.01  $\pm$ 0.09 & 0 & 0.18  $\pm$ 0.07 \\
$4\ell$ (OSSF-0) $S_T$(Mid) && 0 & 0.004  $\pm$ 0.002 & 0 & 0.28  $\pm$ 0.10 & 2 & 2.5  $\pm$ 1.2 \\
$4\ell$ (OSSF-0) $S_T$(Low) && 0 & 0.04  $\pm$ 0.02 & 0 & 2.98  $\pm$ 0.48 & 4 & 3.5  $\pm$ 1.1 \\
$4\ell$ (OSSF-1,~no Z) $S_T$(High) && 1 & 0.009  $\pm$ 0.004 & 0 & 0.10  $\pm$ 0.07 & 0 & 0.12  $\pm$ 0.05 \\
$4\ell$ (OSSF-1,~~Z) $S_T$(High) && 1 & 0.09  $\pm$ 0.01 & 0 & 0.51  $\pm$ 0.15 & 0 & 0.43  $\pm$ 0.15 \\
$4\ell$ (OSSF-1,~no Z) $S_T$(Mid) && 0 & 0.07  $\pm$ 0.02 & 1 & 0.88  $\pm$ 0.26 & 1 & 0.94  $\pm$ 0.29 \\
$4\ell$ (OSSF-1,~~Z) $S_T$(Mid) && 0 & 0.45  $\pm$ 0.11 & 5 & 4.1  $\pm$ 1.2 & 3 & 3.4  $\pm$ 0.9 \\
$4\ell$ (OSSF-1,~no Z) $S_T$(Low) && 0 & 0.09  $\pm$ 0.04 & 7 & 5.5  $\pm$ 2.2 & 19 & 13.7  $\pm$ 6.4 \\
$4\ell$ (OSSF-1,~~Z) $S_T$(Low) && 2 & 0.80  $\pm$ 0.34 & 19 & 17.7  $\pm$ 4.9 & 95 & 60  $\pm$ 31 \\
$4\ell$ (OSSF-2,~no Z) $S_T$(High) && 0 & 0.02  $\pm$ 0.01 & -- & -- & -- & -- \\
$4\ell$ (OSSF-2,~~Z) $S_T$(High) && 0 & 0.89  $\pm$ 0.34 & -- & -- & -- & -- \\
$4\ell$ (OSSF-2,~no Z) $S_T$(Mid) && 0 & 0.20  $\pm$ 0.09 & -- & -- & -- & -- \\
$4\ell$ (OSSF-2,~~Z) $S_T$(Mid) && 3 & 7.9  $\pm$ 3.2 & -- & -- & -- & -- \\
$4\ell$ (OSSF-2,~no Z) $S_T$(Low) && 1 & 2.4  $\pm$ 1.1 & -- & -- & -- & -- \\
$4\ell$ (OSSF-2,~~Z) $S_T$(Low) && 29 & 29  $\pm$ 12 & -- & -- & -- & -- \\
\hline
$3$ Lepton results \\
\hline
$3\ell$ (OSSF-0) $S_T$(High) && 2 & 1.14  $\pm$ 0.43 & 17 & 11.2  $\pm$ 3.2 & 20 & 22.5  $\pm$ 6.1 \\
$3\ell$ (OSSF-0) $S_T$(Mid) && 5 & 7.4  $\pm$ 3.0 & 113 & 97  $\pm$ 31 & 157 & 181  $\pm$ 24 \\
$3\ell$ (OSSF-0) $S_T$(Low) && 17 & 13.5  $\pm$ 4.1 & 522 & 419  $\pm$ 63 & 1631 & 2018  $\pm$ 253 \\
$3\ell$ (OSSF-1,~no Z) $S_T$(High) && 6 & 3.5  $\pm$ 0.9 & 10 & 13.1  $\pm$ 2.3 & -- & -- \\
$3\ell$ (OSSF-1,~~Z) $S_T$(High) && 17 & 18.7  $\pm$ 6.0 & 35 & 39.2  $\pm$ 4.8 & -- & -- \\
$3\ell$ (OSSF-1,~no Z) $S_T$(Mid) && 32 & 25.5  $\pm$ 6.6 & 159 & 141  $\pm$ 27 & -- & -- \\
$3\ell$ (OSSF-1,~~Z) $S_T$(Mid) && 89 & 102  $\pm$ 31 & 441 & 463  $\pm$ 41 & -- & -- \\
$3\ell$ (OSSF-1,~no Z) $S_T$(Low) && 126 & 150  $\pm$ 36 & 3721 & 2983  $\pm$ 418 & -- & -- \\
$3\ell$ (OSSF-1,~~Z) $S_T$(Low) && 727 & 815  $\pm$ 192 & 17631 & 15758  $\pm$ 2452 & -- & -- \\
\hline													\\
Total $4\ell$  && 37 & 42  $\pm$ 13 & 32.0 & 32.1  $\pm$ 5.5 & 124 & 85  $\pm$ 32 \\
Total $3\ell$  & & 1021 & 1137  $\pm$ 198  & 22649 & 19925  $\pm$ 2489  & 1808 & 2222  $\pm$ 255 \\
Total  && 1058 & 1179  $\pm$ 198 & 22681 & 19957  $\pm$ 2489 & 1932 & 2307  $\pm$ 257 \\

\hline \hline
\end{tabular}
\end{center}
\end{table}

In the three-lepton, no-Z channels with low-\jt and low-\met in Table~\ref{tab:ra7} or with low-\ST~in Table~\ref{tab:rpv}, we reject events that have a three-body mass consistent with a Z, which lowers the impact of asymmetric conversions. Because the low-\jt/low-\met bin is not identical to the low-\ST~bin, the two tables have slightly different numbers of events.

The two-$\tau_\mathrm{h}$ selection in Table~\ref{tab:ra7} is chosen to be consistent with Ref.~\cite{Chatrchyan:2011ff}, in which the two reconstructed $\tau_\mathrm{h}$s either both have, or both do not have, an associated $\pi^0$. Improved understanding of $\tau_\mathrm{h}$ reconstruction has allowed us to expand the two-$\tau_\mathrm{h}$ selection in Table~\ref{tab:rpv} to include events where one reconstructed $\tau_\mathrm{h}$ has an associated $\pi^0$ and one does not. This modification provides improved sensitivity in the two-$\tau_\mathrm{h}$ channels.

\begin{figure}
  \begin{center}
  \includegraphics[width=7.5cm]{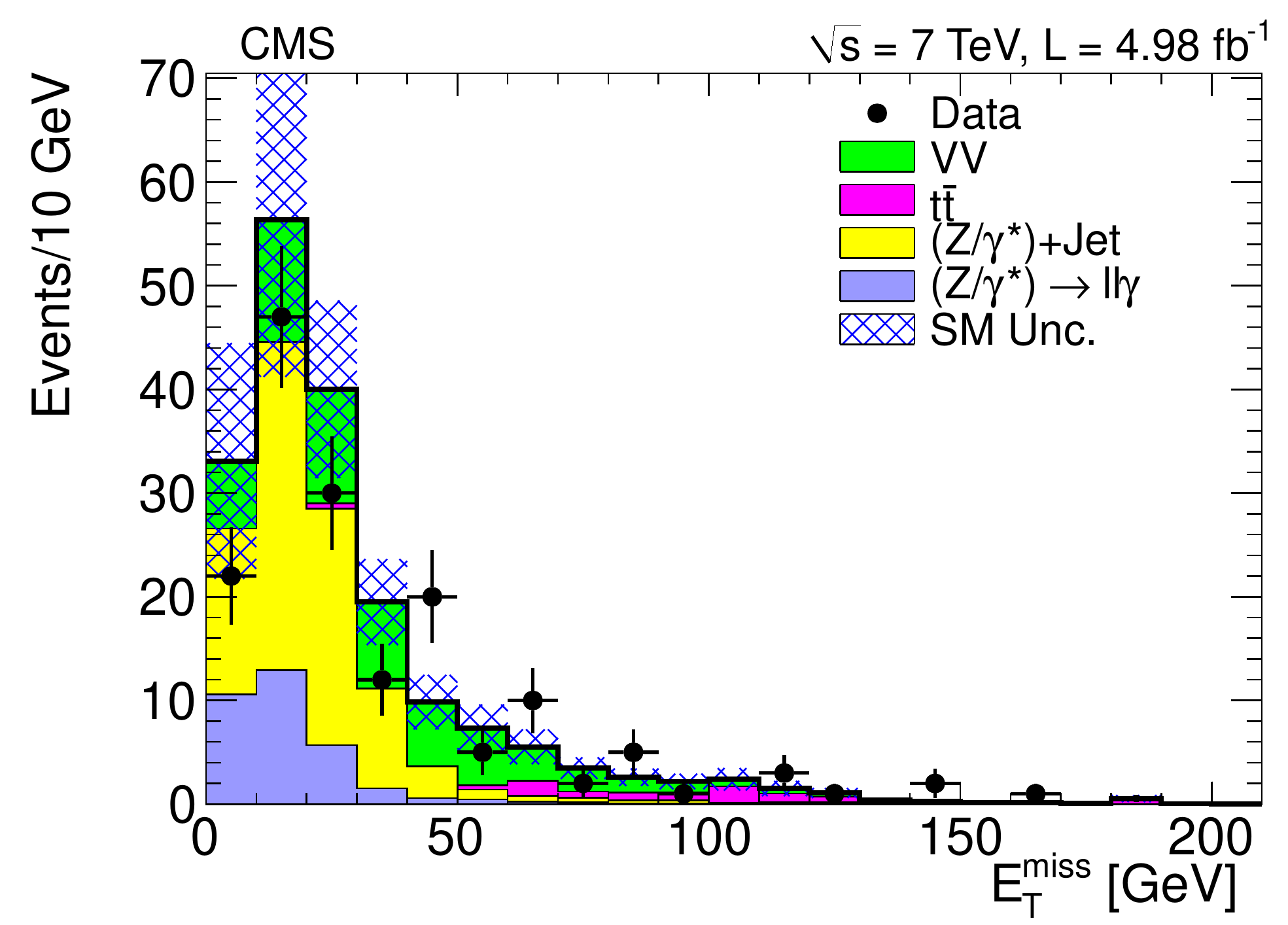}
  \includegraphics[width=7.5cm]{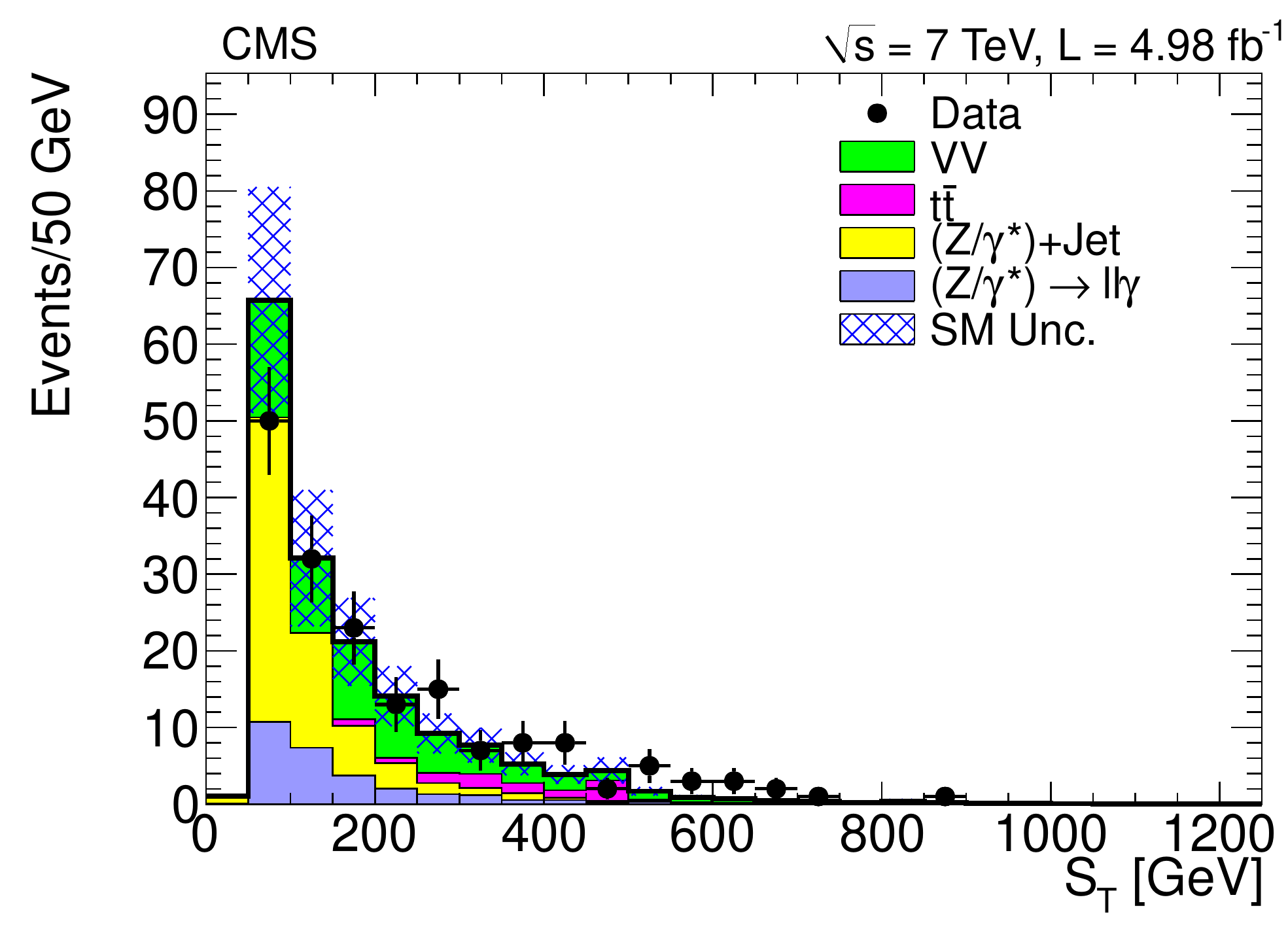}
  \caption{
  \label{fig:Observed_Events_ST}
We show the \met distribution for the three lepton, no-$\tau_\mathrm{h}$, no-Z, $\jt < 200\GeV$ channel (left) and the \ST~distribution for the same set of events (right). Comparison between the observed events (dots) and expected SM background (histograms) is shown. The hashed bands represent the uncertainty on the SM contribution.
}
  \end{center}
 \end{figure}

Tables~\ref{tab:ra7} and~\ref{tab:rpv} illustrate a key feature of this analysis: the division into exclusive channels, some with large SM expectations and some in which they are negligible.  Any specific BSM scenario may produce events in a subset of channels, but not in the rest.  The former constitutes the ``signal'' region for that particular model, while the latter comprises the ``control'' region. The sensitivity of this analysis to a given model depends on the size of the contribution to channels with low SM expectations. Figure~\ref{fig:Observed_Events_ST} shows a representative distribution for each table, for the three-lepton, no-Z signature with zero $\tau_\mathrm{h}$s.

We observe one four-lepton event in the zero-$\tau_\mathrm{h}$, no-Z, high-\met, low-\jt bin in Table~\ref{tab:ra7} and in the high-\ST~bin in Table~\ref{tab:rpv}. The SM expectation for such an event is much lower than one for our dataset. We find that the dominant SM contribution to the bin is from ZZ production, where one of the Z bosons is virtual. The background estimate is obtained with \textsc{MadGraph}~\cite{Alwall:2011uj}, and an uncertainty of 40\% is assigned based on differences in the estimate with \textsc{mcfm}~\cite{Campbell:2011bn}. Consistent predictions in low-\ST~and on-shell control samples of the data are found; however, there are not enough data to test the SM prediction for off-shell diboson production at high \ST.

\section{Interpretation}
In supersymmetry, multilepton final states arise naturally in the subset of GMSB parameter space where the right-handed sleptons are essentially flavor-degenerate and at the bottom of the minimal supersymmetric standard model (MSSM) mass spectrum. Supersymmetric production can proceed through pairs of squarks and gluinos (\PSq\ and \PSg).  Cascade decays of these states eventually pass through the lightest neutralino ($\PSg,\PSq\rightarrow \widetilde{\chi}^0 + X$), which decays into a slepton ($\widetilde{\ell}$) and a lepton ($\chi^0 \rightarrow \widetilde{\ell}^\pm \ell^{\mp}$). Each of the right-handed sleptons promptly decays to the Goldstino component of the almost-massless and non-interacting gravitino and a lepton ($\widetilde{\ell} \to \PXXSG\ell$) thus yielding events with four or more hard leptons and missing transverse energy. Such scenarios have a large cross section with little background~\cite{Alves:2011wf}.

Models with RPV interactions that violate $B$ or $L$, but not both, can avoid direct contradiction with the proton-lifetime upper limits~\cite{Nakamura:2010zzi}.  A common specification of the superpotential includes three RPV terms, parametrized by the Yukawa couplings $\lambda_{ijk}$, $\lambda_{ijk}^\prime$ or $\lambda_{ijk}^{\prime \prime}$~\cite{Barbier:2004ez}, respectively,

$$W_{R\mathrm{PV}} \,=\, \frac{1}{2}\,
\lambda_{ijk}L_iL_j\overline{E}_k\,+\,
\lambda^\prime_{ijk}L_iQ_j\overline{D}_k\,+\,
\frac{1}{2}\, \lambda^{\prime \prime}_{ijk}\overline{U}_i\overline{D}_j\overline{D}_k,$$

where $i,j,$ and $k$ are generation indices; $L$ and $Q$ are the lepton and quark $SU(2)_L$ doublet superfields; and $\overline{E}$, $\overline{D}$, and $\overline{U}$ are the charged lepton, down-like quark, and up-like quark $SU(2)_L$ singlet superfields, respectively.  The third term violates baryon-number conservation, while the first and second terms are lepton-number violating. In this analysis, we consider leptonic $R$-parity-violating (L-RPV) models with $\lambda_{ijk}\ne0$ and $\lambda^\prime_{ijk} = \lambda^{\prime\prime}_{ijk} =0$, as well as hadronic $R$-parity-violating (H-RPV) models with  $\lambda_{ijk} = \lambda^{\prime}_{ijk} = 0 $  and $ \lambda^{\prime\prime}_{ijk} \ne 0$.  We consider squark and gluino production with leptons coming either from the decay of a neutralino through leptonic RPV (L-RPV) couplings, or from cascade decays to a neutralino that decays through hadronic RPV couplings (H-RPV). The value of $\lambda_{ijk}$ determines the lifetime and therefore the decay length of the intermediate particle. Values of $\lambda_{ijk}$ and $\lambda^{\prime\prime}_{ijk}$ considered in this analysis, 0.05, correspond to decay lengths less than 100\mum, which is chosen so that most decays will be prompt.

The files specifying the signal-model parameters are generated according to the SUSY Les Houches accord (SLHA) standards with the ISAJET program~\cite{Skands:2003cj,Baer:1993ae}. The SLHA output files are input to {\sc Pythia} for event generation using the CTEQ6.6 PDFs. The generated events then undergo detector simulation in the CMS fast simulation framework~\cite{Abdullin:1328345}. The cross sections are calculated in {\sc Pythia} to leading order with NLO corrections calculated using {\sc Prospino}~\cite{Beenakker:1996ed}.

Simulation for the co-NLSP scenario is generated on a grid in the chargino-gluino mass plane. The other super partner masses are related to these by $m_{\widetilde{\ell}_R} = 0.3 m_{\chi^{\pm}}$, $m_{\widetilde{\chi}_1^0} = 0.5 m_{\chi^{\pm}}$, $m_{\widetilde{\ell}_L} = 0.8 m_{\chi^{\pm}}$, and $m_{\PSq} = 0.8 m_{\PSg}$. Flavor universality and vanishing left-right mixing for squarks and sleptons are enforced. Simulations for three separate L-RPV models and the H-RPV model, described below, are generated on a grid in the squark--gluino mass plane. To determine the sensitivity for various signal-model scenarios, we perform a simultaneous fit across all of the exclusive channels listed in either Table~\ref{tab:ra7} or in Table~\ref{tab:rpv} to compute the likelihood of observing a signal.

We present the observed limits, the median expected limits, and the 1- and 2-standard deviation bands at each point in the mass planes of the models of interest, which are calculated using the ``LHC style"~\cite{ATLAS:1379837} CLs~\cite{Junk:1999kv} prescription. The inputs to the limit calculation include the number of observed events and background estimates as listed in Tables~\ref{tab:ra7} and~\ref{tab:rpv}, and signal estimates obtained for the model point. The systematic and statistical uncertainties on the signal and background estimates are treated as nuisance parameters in the limit calculations, with appropriate correlations taken into account. We estimate the effect of uncertainties from the PDFs as 14\% and from scale uncertainty as 10\% on the signal strengths.

\begin{figure}
\begin{center}
\includegraphics*[width=7.5cm]{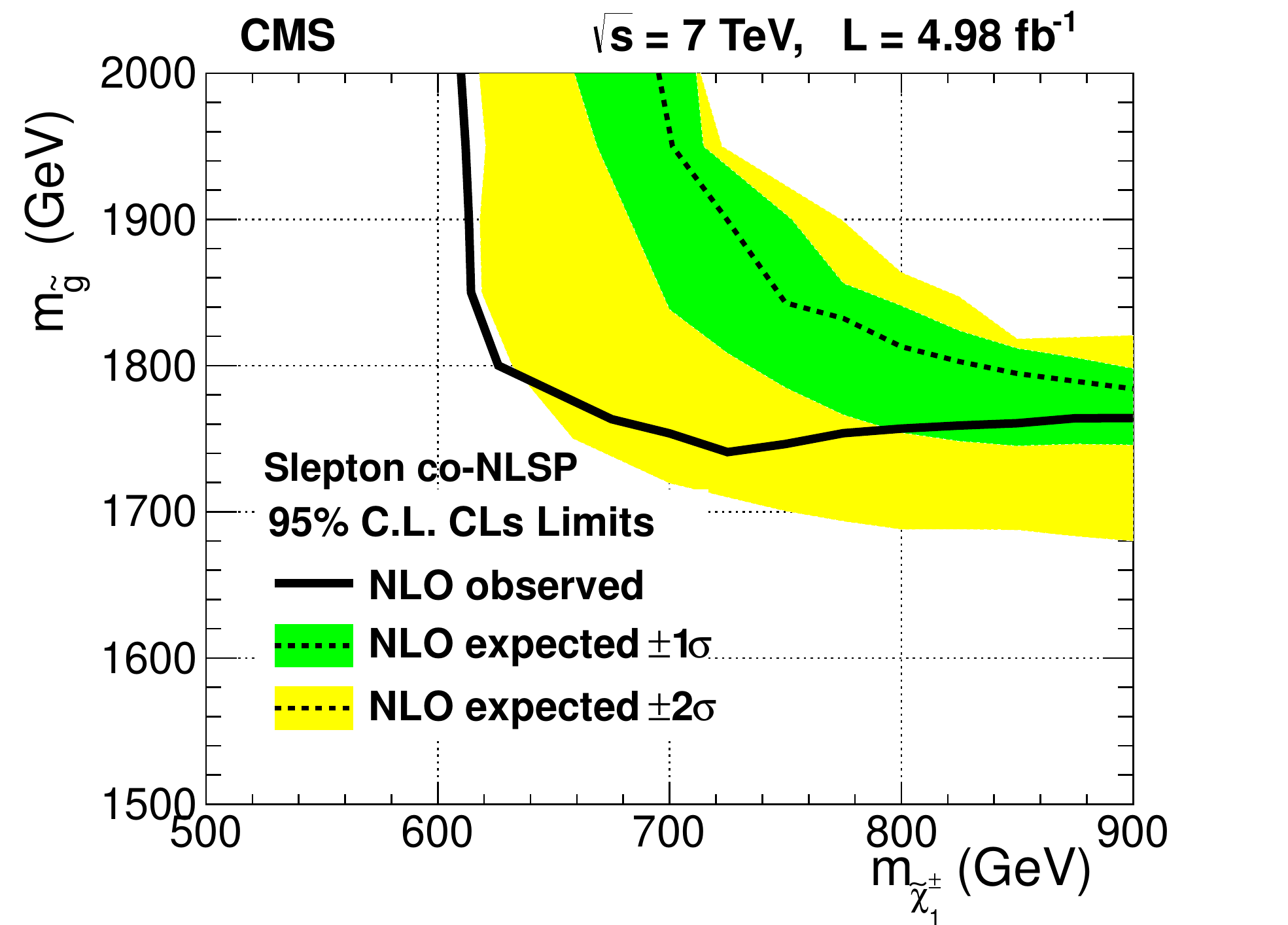}
\caption{Excluded region in the gluino mass versus wino-chargino mass plane for the slepton co-NLSP scenario described in the text. The region below the solid, black line (observed limit) is excluded at 95\% CL.  For comparison, the expected limits are shown as well. The deviation of the observed curve from the expected curve is driven by the four-lepton, one-$\tau_\mathrm{h}$, $\met>50$ \GeV, $\jt<200$ \GeV, no Z channel, in which we observed a slightly larger number of events than the expectation.
 }
\label{fig:limit1}
\end{center}
\end{figure}

We interpret the \jt/\met~binning (Table~\ref{tab:ra7}) in the co-NLSP model. The 95\% confidence level (CL) exclusion limits for the slepton co-NLSP model are shown in the chargino-gluino mass plane in  Fig.~\ref{fig:limit1}. The exclusion curve approaches a horizontal asymptote in regions dominated by strong superpartner production, and the vertical one in regions dominated by weak superpartner production. With strong superpartners decoupled, the production is dominated by wino chargino-neutralino and chargino-chargino production.

\begin{figure}[htbp]
\centering
\begin{tabular}{cc}

\includegraphics*[width=7.5cm]{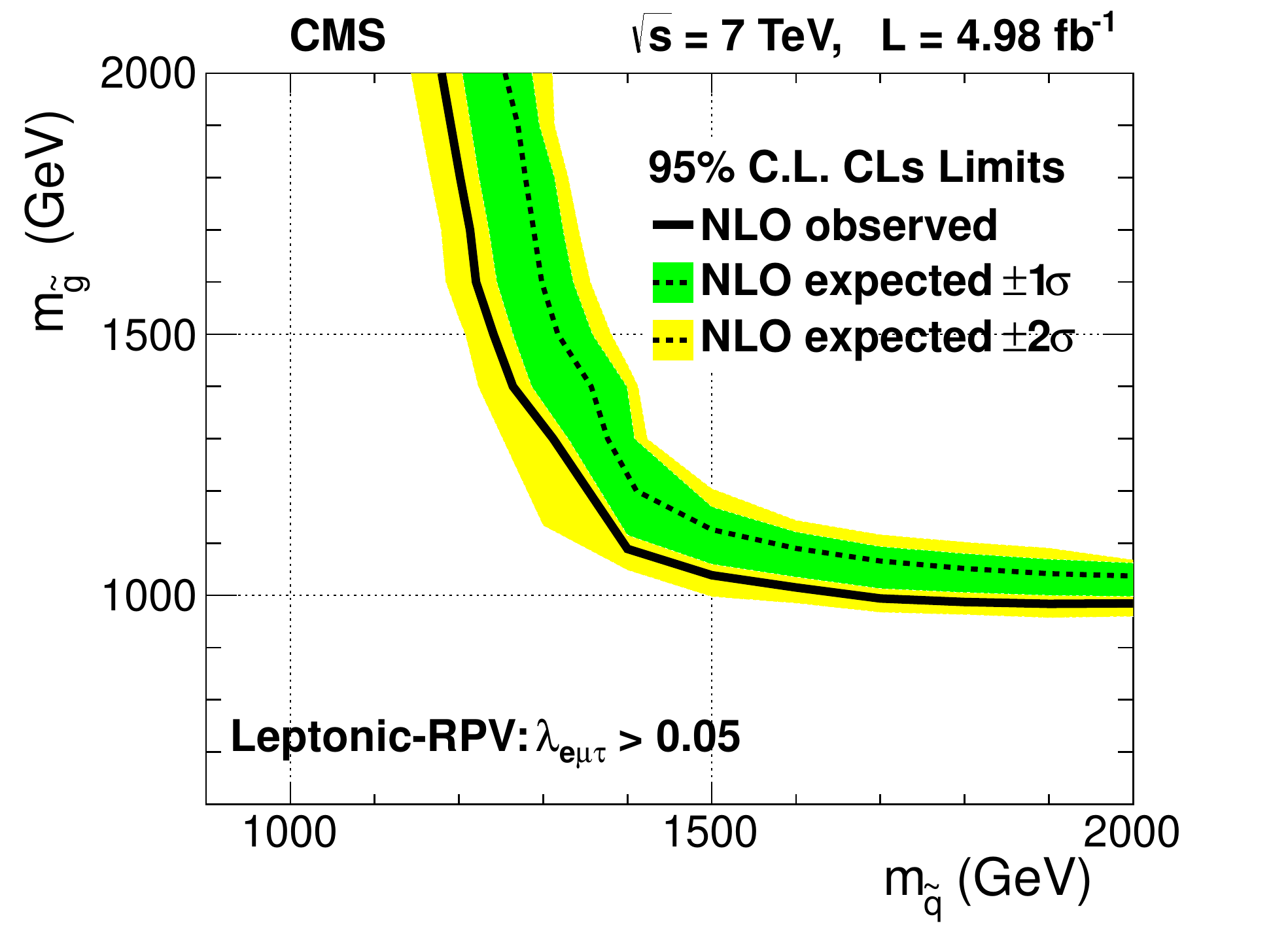}
\includegraphics*[width=7.5cm]{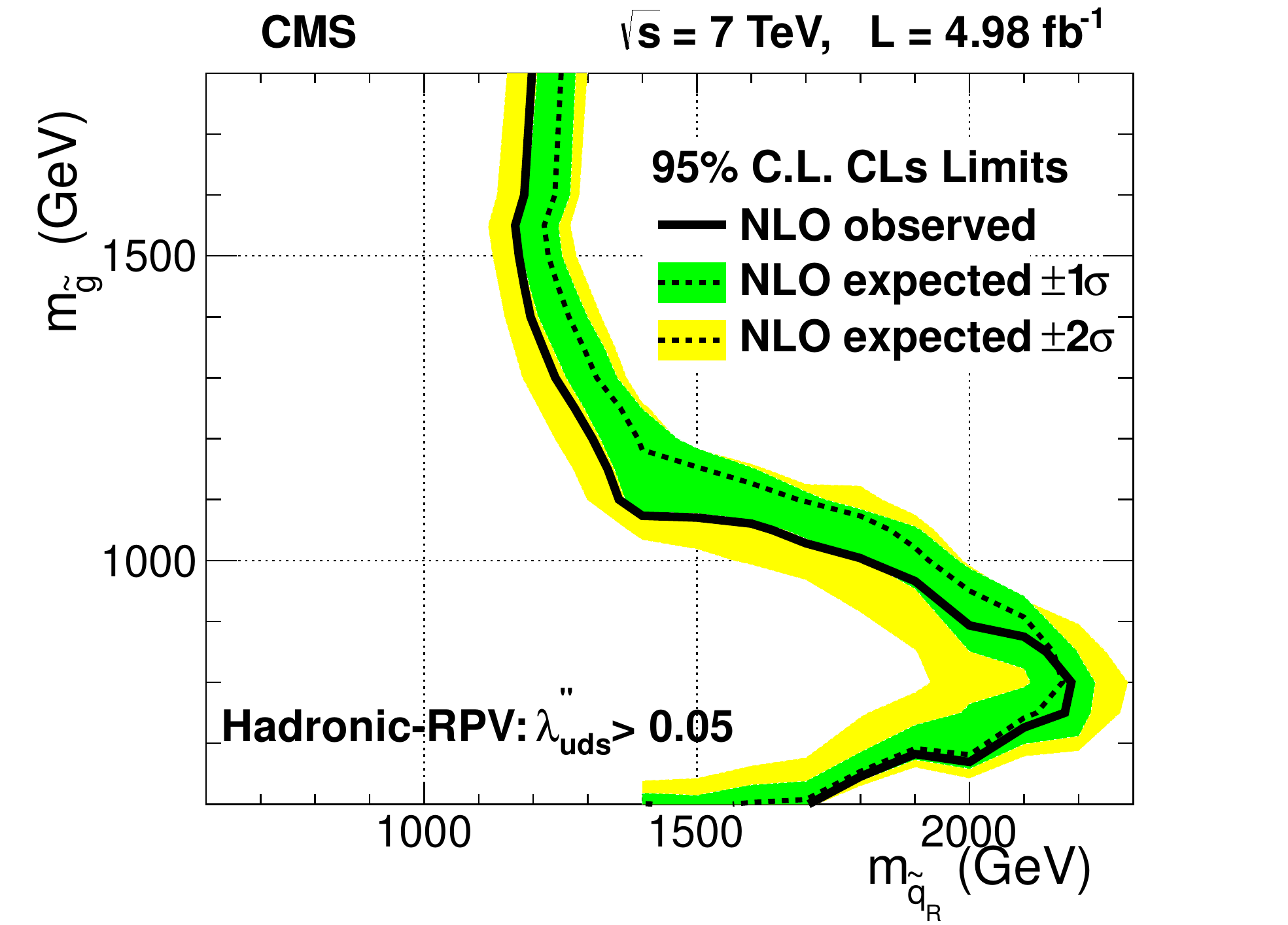}

\end{tabular}
\caption{
   \label{fig:limits_with_expected}
The 95\% CL exclusion regions in the squark and gluino mass plane for the model with the RPV coupling $\lambda_{e\mu\tau} > 0.05$ (left) and the H-RPV $\lambda_{uds}^{''} > 0.05$ (right) . The observed limits, along with limits expected in the absence of signal, are shown along with the uncertainty in the expectation. The regions to the left of the curves are excluded.
}
\end{figure}

The \ST~binning from Table~\ref{tab:rpv} is interpreted in the context of $R$-parity violating models. In Fig.~\ref{fig:limits_with_expected}, we show the 95\% CL exclusion-limit contours for $\lambda_{{\Pe}\mu\tau}$ and H-RPV coupling $\lambda_{uds}^{\prime\prime}$ in the squark-gluino mass space, along with the expected limits in the absence of signal.

In the specific slepton co-NLSP L-RPV SUSY topology described in Ref.~\cite{Chatrchyan:2011ff} and references therein, the bino is the lightest superpartner with a fixed mass of 300\GeV.  The gluino and degenerate squark masses, $m_{\PSg}$ and $m_{\PSq}$, are variable and define the parameter space for our search.  All other superpartners are decoupled, holding the bino RPV decay width fixed.

The superpartner spectrum for the H-RPV SUSY topology used here consists of a wino, right-handed sleptons, and bino, with fixed masses of 150, 300, and 500\GeV, respectively, and varying gluino and right-handed squark masses larger than 500\GeV. The left-handed squark masses and higgsino mass parameter are fixed at 5000 and 3000\GeV, respectively. Flavor universality and vanishing left-right mixing for squarks and sleptons are enforced.

In this topology, the right-handed squarks decay to the bino and the gluino decays predominantly to the bino except for relatively small values of the gluino--bino mass splitting. The bino decays to a right-handed slepton, which in turn decays to the wino neutralino. Starting from strongly-interacting superpartner pair production, all cascade decays that produce the bino therefore yield either four leptons, of which zero, two, or four can be taus. The wino lightest superpartner decays to three jets through hadronic R-parity violating couplings. This topology yields events with jets and multiple charged leptons, with no particles emitted directly from the supersymmetric cascade that carry missing energy.

In the H-RPV case, gluino masses below 500\GeV are not excluded even though the production cross section in this region can be large. This is due to the low gluino branching fraction to the bino and subsequently to leptons.  The non-zero coupling is $\lambda_{\cPqu\cPqd\cPqs}^{\prime\prime}$ in our H-RPV model. We apply our search findings to RPV models in which either $\lambda_{\Pe\mu\mu}$, $\lambda_{\Pe\mu\tau}$, or $\lambda_{\mu\tau\tau}$ couplings are non-zero, though only the $\lambda_{\Pe\mu\tau}$ results are shown here due to space constraints. We choose $\lambda_{{\Pe}\mu\tau}$ because it couples democratically to all three types of leptons, although the sensitivity is typically better for couplings to only electrons and muons.


\section{Conclusions}
We have performed a search for physics beyond the standard model by examining a variety of multilepton final states. By studying many channels with different requirements, we greatly enhance sensitivity to new physics. We see good agreement between observations and expectations in all exclusive channels, both in channels with and without Z-boson decays.

Taking advantage of the 7\TeV center-of-mass energy at the LHC, we are able to probe new regions of the MSSM parameter space. Our search complements those at the Tevatron~\cite{Aaltonen:2008pv,Abulencia:2007mp,Abazov:2006nw}, which are mostly sensitive to electroweak gaugino production via quark-antiquark interactions~\cite{Chatrchyan:2011ff}. The results presented here are mostly sensitive to gluino and squark production via quark-gluon or gluon-gluon interactions. Finding consistency with SM expectations, we use these results to exclude regions of slepton co-NLSP scenarios with gravitinos as the LSP as described above.

We demonstrate the reach and versatility of the search by applying the results to the case of RPV decays of SUSY particles in multilepton events.  We are able to exclude squark and gluino masses in the 1\TeV range for models with a neutralino LSP that decays through a L-RPV coupling $\lambda_{\Pe\mu\tau}$ that is greater than 0.05. Similarly, we are able to exclude regions of a model with leptons emitted in cascade decays without missing energy and a neutralino LSP that decays through the H-RPV coupling $\lambda_{\cPqu\cPqd\cPqs}^{\prime\prime} > 0.05$.  

\section*{Acknowledgements}

\hyphenation{Bundes-ministerium Forschungs-gemeinschaft Forschungs-zentren} We wish to congratulate our colleagues in the CERN accelerator departments for the excellent performance of the LHC machine. We thank the technical and administrative staff at CERN and other CMS institutes. This work was supported by the Austrian Federal Ministry of Science and Research; the Belgium Fonds de la Recherche Scientifique, and Fonds voor Wetenschappelijk Onderzoek; the Brazilian Funding Agencies (CNPq, CAPES, FAPERJ, and FAPESP); the Bulgarian Ministry of Education and Science; CERN; the Chinese Academy of Sciences, Ministry of Science and Technology, and National Natural Science Foundation of China; the Colombian Funding Agency (COLCIENCIAS); the Croatian Ministry of Science, Education and Sport; the Research Promotion Foundation, Cyprus; the Estonian Academy of Sciences and NICPB; the Academy of Finland, Finnish Ministry of Education and Culture, and Helsinki Institute of Physics; the Institut National de Physique Nucl\'eaire et de Physique des Particules~/~CNRS, and Commissariat \`a l'\'Energie Atomique et aux \'Energies Alternatives~/~CEA, France; the Bundesministerium f\"ur Bildung und Forschung, Deutsche Forschungsgemeinschaft, and Helmholtz-Gemeinschaft Deutscher Forschungszentren, Germany; the General Secretariat for Research and Technology, Greece; the National Scientific Research Foundation, and National Office for Research and Technology, Hungary; the Department of Atomic Energy and the Department of Science and Technology, India; the Institute for Studies in Theoretical Physics and Mathematics, Iran; the Science Foundation, Ireland; the Istituto Nazionale di Fisica Nucleare, Italy; the Korean Ministry of Education, Science and Technology and the World Class University program of NRF, Korea; the Lithuanian Academy of Sciences; the Mexican Funding Agencies (CINVESTAV, CONACYT, SEP, and UASLP-FAI); the Ministry of Science and Innovation, New Zealand; the Pakistan Atomic Energy Commission; the State Commission for Scientific Research, Poland; the Funda\c{c}\~ao para a Ci\^encia e a Tecnologia, Portugal; JINR (Armenia, Belarus, Georgia, Ukraine, Uzbekistan); the Ministry of Education and Science of the Russian Federation, the Federal Agency of Atomic Energy of the Russian Federation, Russian Academy of Sciences, and the Russian Foundation for Basic Research; the Ministry of Science and Technological Development of Serbia; the Ministerio de Ciencia e Innovaci\'on, and Programa Consolider-Ingenio 2010, Spain; the Swiss Funding Agencies (ETH Board, ETH Zurich, PSI, SNF, UniZH, Canton Zurich, and SER); the National Science Council, Taipei; the Scientific and Technical Research Council of Turkey, and Turkish Atomic Energy Authority; the Science and Technology Facilities Council, UK; the US Department of Energy, and the US National Science Foundation.

Individuals have received support from the Marie-Curie programme and the European Research Council (European Union); the Leventis Foundation; the A. P. Sloan Foundation; the Alexander von Humboldt Foundation; the Belgian Federal Science Policy Office; the Fonds pour la Formation \`a la Recherche dans l'Industrie et dans l'Agriculture (FRIA-Belgium); the Agentschap voor Innovatie door Wetenschap en Technologie (IWT-Belgium); and the Council of Science and Industrial Research, India.

\bibliography{auto_generated}   

\cleardoublepage \appendix\section{The CMS Collaboration \label{app:collab}}\begin{sloppypar}\hyphenpenalty=5000\widowpenalty=500\clubpenalty=5000\textbf{Yerevan Physics Institute,  Yerevan,  Armenia}\\*[0pt]
S.~Chatrchyan, V.~Khachatryan, A.M.~Sirunyan, A.~Tumasyan
\vskip\cmsinstskip
\textbf{Institut f\"{u}r Hochenergiephysik der OeAW,  Wien,  Austria}\\*[0pt]
W.~Adam, T.~Bergauer, M.~Dragicevic, J.~Er\"{o}, C.~Fabjan, M.~Friedl, R.~Fr\"{u}hwirth, V.M.~Ghete, J.~Hammer\cmsAuthorMark{1}, N.~H\"{o}rmann, J.~Hrubec, M.~Jeitler, W.~Kiesenhofer, V.~Kn\"{u}nz, M.~Krammer, D.~Liko, I.~Mikulec, M.~Pernicka$^{\textrm{\dag}}$, B.~Rahbaran, C.~Rohringer, H.~Rohringer, R.~Sch\"{o}fbeck, J.~Strauss, A.~Taurok, P.~Wagner, W.~Waltenberger, G.~Walzel, E.~Widl, C.-E.~Wulz
% ...
\vskip\cmsinstskip
\dag:~Deceased\\
1:~~Also at CERN, European Organization for Nuclear Research, Geneva, Switzerland\\
% ...

\end{sloppypar}
\end{document}